\documentclass[pra,preprint,showpacs]{revtex4-1}

\usepackage{graphicx}

\usepackage{color}

\begin{document}


\title{Origin of the low energy resonance in the double photoionization
of pyrene and coronene, and its absence in the double photoionization of
corannulene}



\author{Ralf Wehlitz}
\email[]{rwehlitz@gmail.com}
\affiliation{Physics Department, University of 
Wisconsin--Madison, Madison, Wisconsin 53706, USA}


\author{David L. Huber}
\email[]{dhuber@wisc.edu}
\affiliation{Physics Department, University of 
Wisconsin--Madison, Madison, Wisconsin 53706, USA}


\date{\today}

\begin{abstract}
The low energy resonance in the double photoionization of the aromatic 
hydrocarbons pyrene (C$_{16}$H$_{10}$) and coronene (C$_{24}$H$_{12}$)
is investigated theoretically using an 
approach based on the one-dimension Hubbard model for $\pi$-conjugated 
systems with nearest-neighbor interactions.  The Independent Subsystem 
Approximation, where the perimeter and interior carbon atoms are treated 
as independent entities, is employed. Since no low energy resonances have 
been found in aromatic hydrocarbons where there are only perimeter carbon 
atoms, we attribute the low energy resonances in pyrene and coronene to the 
interior carbon atoms. However, corannulene (C$_{20}$H$_{10}$) having five 
interior carbon atoms does not exhibit a low-energy resonance in the 
experimental data. We attribute the absence of this resonance to the odd 
number of interior carbon atoms.
\end{abstract}

\pacs{33.80.Eh} 

\maketitle 

\section{INTRODUCTION}
\label{intro}
The double-photoionization (DPI) process, i.e., the removal of two electrons 
by a single photon, has been investigated for various aromatic molecules 
using synchrotron radiation in recent years.  A convenient quantity to 
study the DPI process in atoms and molecules is the 
ratio of doubly to singly charged ions as a function of photon 
energy.\cite{Wehl10}
Common to all atoms and molecules is the knock-out mechanism\cite{Schn02}
in which the 
photoionized electron knocks out a second electron, leading to a 
helium-like ratio curve. In the case of molecules, it has been discovered 
that additional photoionization mechanisms, not observed for atoms, can 
contribute to the production of doubly charged molecular 
ions.\cite{Wehl16}

For aromatic molecules with 5-member rings (pyrole, furan, 
selenophene)\cite{Hart13a} and for the aromatic molecule with the 7-member 
ring (tropone)\cite{Hart17} the ratio curves show a linear increase in 
the ratio above a certain 
threshold, which is far above the double-ionization threshold. It has 
been proposed that the linear increase is associated with the 
simultaneous emission of two electrons with equal kinetic energy and 
opposite momenta.\cite{Hube14,Hube19}

In the case of the aromatic molecules with 6-member rings (e.g. 
pyrrole, benzene, halogenated benzene, pentacene, pyridine, triphenylene, 
coronene) an additional broad peak in the ratio is present.  For 
these molecules, the broad peak is essentially at the same value of the 
energy and is only slightly changed by introducing nitrogen atoms into 
the ring (pyridine, pyrimidine, 1,3,5-triazine)\cite{Wehl15}
or by replacing hydrogen atoms with halogen atoms 
(C$_6$BrF$_5$).\cite{Mash16}

There are two explanations for the origin of this peak. In the first, 
the peak is associated with the break-up of a Coulomb-pair whose formation 
was studied by Mahajan and Thyagaraga.\cite{Maha06}
In the second explanation, the 
peak is associated with the formation of a ``Cooper pair'' whose de Broglie 
wavelength approximately matches the carbon--carbon distance of the 
molecule leading to a standing wave in the $\pi$ orbital.\cite{Wehl12}
It is worthwhile to mention that the $\pi$ orbital is essential for 
this feature.  An experiment on tribromoborazine [(BrBNH)$_3$], a molecule 
with a 6-member ring of boron and nitrogen atoms, does not show any peak 
in its ratio curve but only the contribution of the knock-out mechanism and 
a linear increase of the ratio.\cite{Wehl17}

Recently it has been shown that the DPI curves of chlorobenzene and 
fluorobenzene have a resonance at the onset of the molecular DPI near 
4 eV.\cite{Sun21} This resonance has been interpreted in their paper as 
being due to the de Broglie wavelength of the two-electron pair 
matching the size of the conjugated p--$\pi$ orbital.

\begin{figure}[htb]
\includegraphics[width=8.0cm,clip]{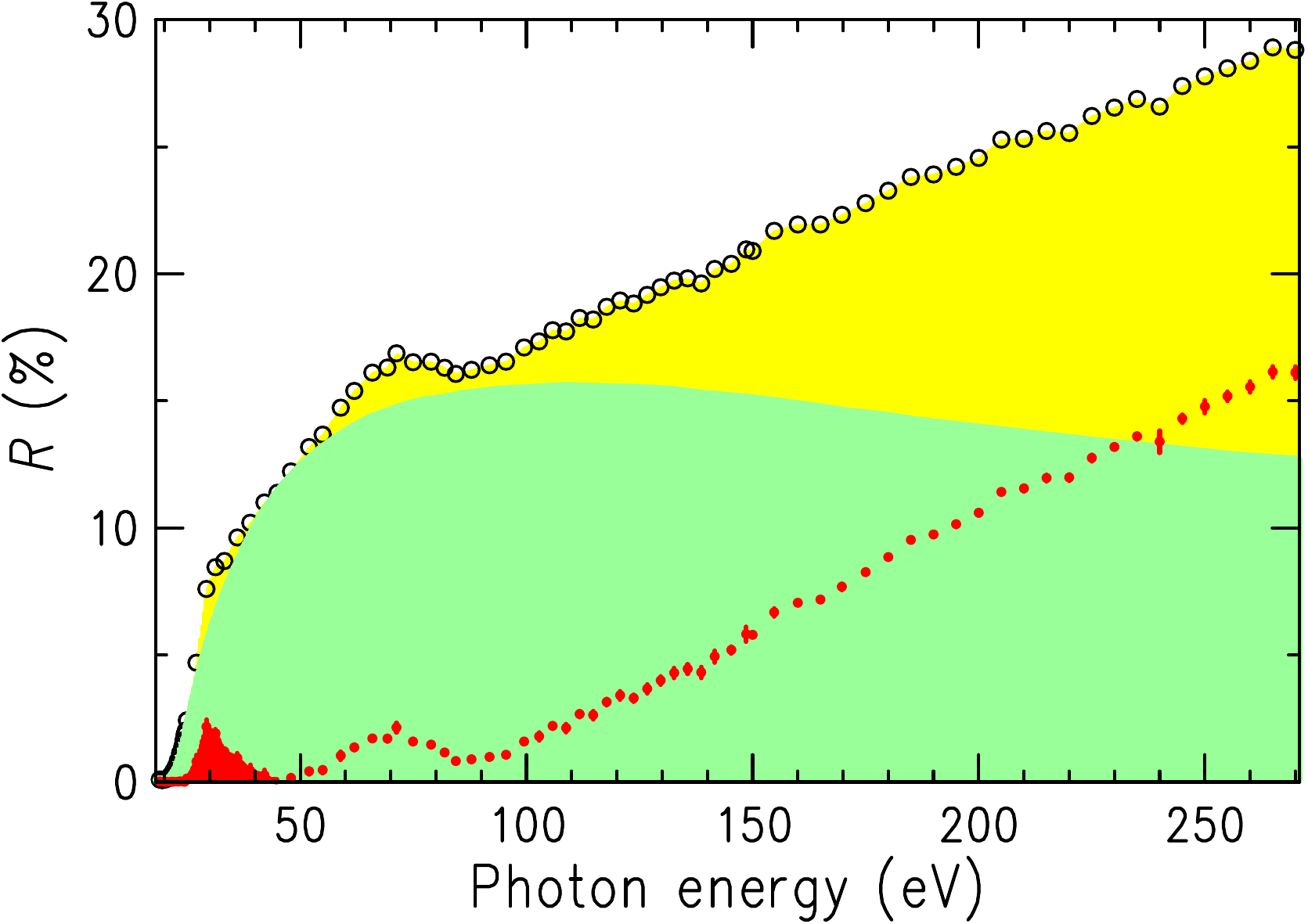} 
\caption{\small Extended plot of the DPI ratio 
$R = M^{2+} / (M^{1+} + M^{2+})$ of pyrene as a function of photon energy 
(open circles).\cite{Wehl14a} $M^{1+}$ and $M^{2+}$ are the yields of 
singly and doubly charged parent ions, respectively.
The dark (green) shaded area represents the contribution from the knock-out 
mechanism. The red (filled) circles represent the ratio after subtracting the 
knock-out contribution, which rises linearly above 95 eV .
}
\label{allpyrene}
\end{figure}

\begin{figure}[htb]
\includegraphics[width=7.0cm,clip]{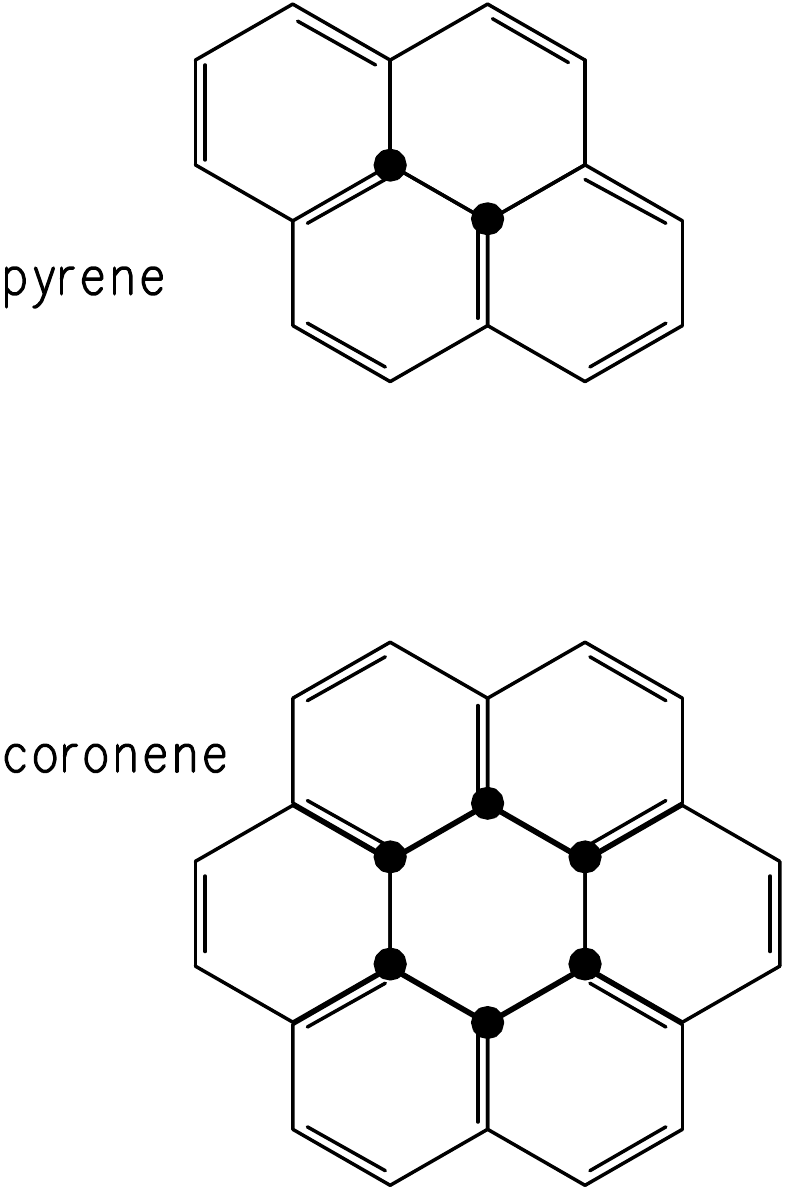} 
\caption{\small Location of the interior carbon atoms in pyrene and 
coronene indicated by solid circles.
}
\label{mol}
\end{figure}

Two of the molecules investigated here (pyrene\cite{Wehl14a} and 
coronene\cite{Wehl12}) exhibit peaks in 
their ratios about 10 eV above the DPI threshold. These peaks are distinct 
from the resonances studied in Ref.\ \onlinecite{Sun21} which we did not 
observe in our studies probably due to the larger energy-step size. 
Figure \ref{allpyrene} illustrates the effect of 
the different mechanisms contributing to the production of doubly charged 
parent ions of pyrene. In contrast to the molecules mentioned previously, 
pyrene and coronene have interior carbon atoms, i.e., atoms that are not 
located on the outer edge as shown in Fig.\ \ref{mol}. In this paper, we 
will discuss a possible connection between the 10-eV peak and the 
interior carbon atoms. Section \ref{analysis} describes the theoretical 
model used in our analysis, Section \ref{pyrene} and Section \ref{coronene} 
applies our model to pyrene and coronene, respectively. We discuss the 
absence of a 10-eV peak for corannulene in Section 
\ref{corannulene}. Our interpretation is summarized in Section \ref{summary}.

\section{THEORETICAL ANALYSIS OF LOW-ENERGY PEAK}
\label{analysis}
The experimental results presented here are plots of the DPI ratio $R$ 
versus excess energy relative to the DPI threshold. $R$ is defined by the 
equation
\begin{equation}
R = M^{2+} / M_{tot} - K
\label{one}
\end{equation}
in which $M^{2+}$ and $M_{tot}$ denote the yields of doubly charged and 
singly plus doubly charged parent ions, respectively.  The contribution from
triply charged ions is assumed to be small and is not included in 
$M_{tot}$. $K$ denotes the 
knock-out contribution to the DPI process modeled by a scaled helium 
ratio curve.   

In analyzing the low energy peaks near 10 eV, we make use of the Independent 
Subsystem Approximation, in which the perimeter and interior carbon 
atoms are treated as separate systems.\cite{Hube14,Hubx19}
The low energy resonance in 
pyrene is attributed to the interaction between the two $\pi$ electrons 
associated with the two interior carbon atoms. To understand this 
point, we focus our analysis on the 2-site Hubbard model.  In one 
dimension, the Hubbard model is equivalent to the Pariser-Parr-Pople 
model of $\pi$-conjugated systems where the 2-site Hamiltonian has the 
form\cite{Burs98}
\begin{equation}
H = -t (c_1^*c_2 + c_1c_2^*) + U
\label{two}
\end{equation}
in which $t$ denotes the electron transfer integral, $U$ is the 
interaction between two electrons occupying the same site and $c$ and 
$c^*$ denote electron annihilation and creation operators, respectively.

\section{PYRENE}
\label{pyrene}
An analysis of the photoexcited states in dimerized Mott insulators has 
been outlined by N.\ Maeshima and K.\ Yonemitsu.\cite{Maes06}
Equation (20) and related text in Section Vc of Ref.\ \onlinecite{Maes06} 
are directly applicable to the 2-site Hubbard model.  The excitation 
energy of the model associated with the optical absorption, $E_0$, 
is expressed as 
\begin{equation}
E_0 = U/2 + (U^2/4 + 4t^2)^{1/2}.
\label{three}
\end{equation}
Note that there is only a single excitation energy for the two-electron 
system. In our analysis of the DPI in pyrene, we identify the energy of 
the resonance peak relative to threshold with $E_0$.  To obtain a 
theoretical estimate of $E_0$, we need values of both $U$ and $t$.  
Values for $U$ and $t$ reported in the benzene literature are 10.06 eV 
and 2.54 eV, respectively.\cite{Burs98}  
The parameter $U$ is associated with a single carbon atom. As such, it is 
expected to be nearly the same for all carbon atoms in an aromatic 
molecule. In contrast, the parameter $t$ is associated with a pair
of carbon atoms and will depend on the details of the electron transfer 
between the two, which can vary with the location of the pair.

\begin{figure}[htb]
\includegraphics[width=8.0cm, clip]{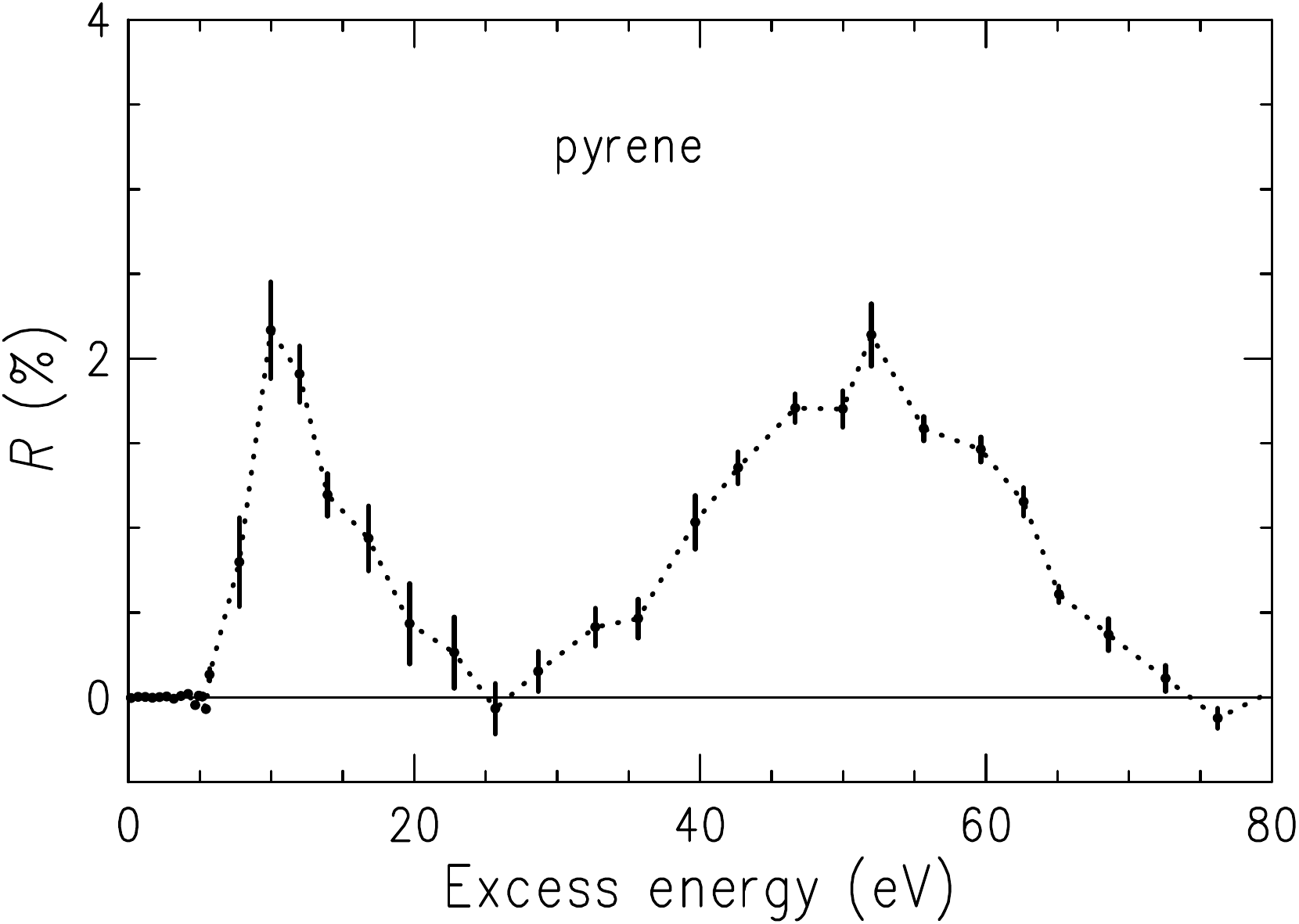}
\caption{\small DPI ratio $R$ vs.\ excess energy relative to the DPI
threshold for pyrene.\cite{Wehl14a} $R$ is defined in Eq.\ (\ref{one}). 
}
\label{pyr}
\end{figure}

The resonance in the DPI of pyrene is attributed 
to the absorption of a photon by the electron pair inside the 
perimeter of the molecule at the interior caron atoms (cf.\ Fig.\ \ref{mol}).
This process promotes the molecule to an excited state. The molecule 
subsequently ionizes creating a pair of free electrons with total 
kinetic energy $E_0$. The experimental energy of the resonance $E_0$ 
in the DPI of pyrene shown in Fig.\ \ref{pyr} is equal to ($10.5 \pm 1.0$) eV. 

\begin{figure}[htb]
\includegraphics[width=8.0cm, clip]{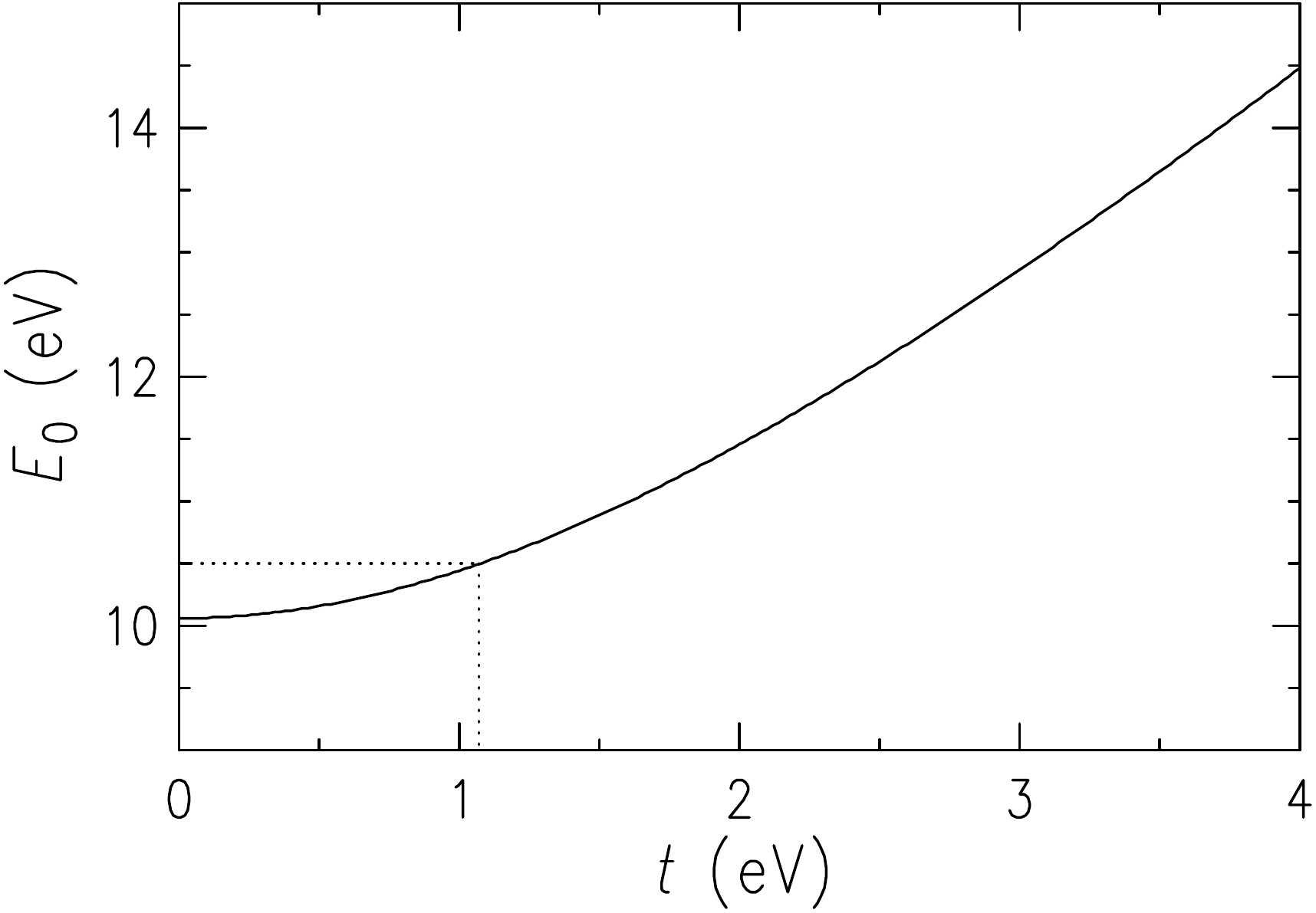}
\caption{\small Excitation energy $E_0$ as a function of the transfer
integral $t$ for a fixed value $U=10.06$ eV using Eq.\ (\ref{three}).
The dotted line starting at $E_0$=10.5 eV corresponds to a $t$ of 1.07 eV}.
\label{fignew}
\end{figure}

An extension of the analysis, which is analogous to an approach followed 
in Ref.\ \onlinecite{Hube19}, is to keep $U$ fixed at the benzene value, 
10.06 eV. The value of $t$ is then adjusted to give the experimental energy of 
the peak, 10.5 eV. As shown in Fig.\ \ref{fignew} the resulting value of 
$t$ is 1.07 eV, indicating that the
value of the transfer integral of the two interior carbon atoms in pyrene
is approximately $1/3$ of the benzene value of 2.54 eV.

\section{CORONENE}
\label{coronene}
\begin{figure}[htb]
\includegraphics[width=8.0cm,clip]{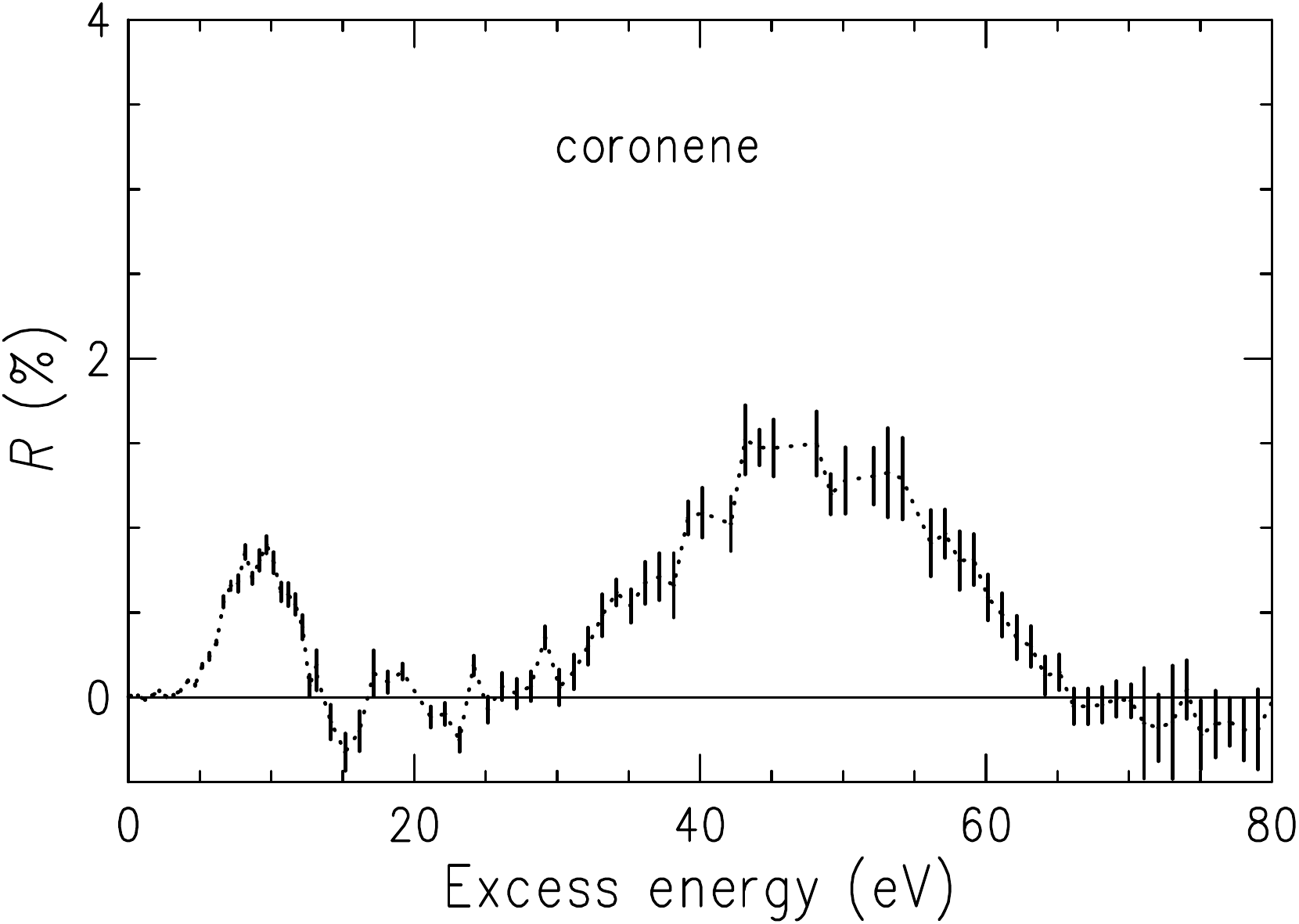}
\caption{\small DPI ratio $R$ vs.\ excess energy relative to the DPI 
threshold for coronene.\cite{Wehl12} $R$ is defined in Eq.\ (\ref{one}).
}
\label{four}
\end{figure}

In coronene, there are six interior carbon atoms arranged in a 
benzene-like ring (cf.\ Fig.\ \ref{mol}). In the Independent Subsystem 
Approximation, the interior atoms are treated as a separate system. 
In pyrene as well as coronene, each interior carbon atom has three nearest 
neighbors.  The two molecules differ in that the interior carbon atom in 
pyrene has one nearest-neighbor that is also an interior atom, whereas in 
coronene, the interior atoms have two interior nearest-neighbors.
Therefore, the 2-site Hubbard model used for pyrene can, unfortunately, 
not be applied to coronene.

The experimental DPI data in Fig.\ \ref{four} are evidence of a well-defined 
resonance at ($9.4 \pm 1.0$) eV.
The similarity of the profile of the 9.4-eV resonance in coronene to
the profile of the 45-eV resonance in coronene and benzene suggests
a common origin: the break-up of a closed-loop de Broglie wave of a
two-electron pair.
We note that we attribute the fluctuations in the ratio around 20 eV
to experimental uncertainties.

\section{CORANNULENE}
\label{corannulene}
\begin{figure}[htb]
\includegraphics[width=8.0cm,clip]{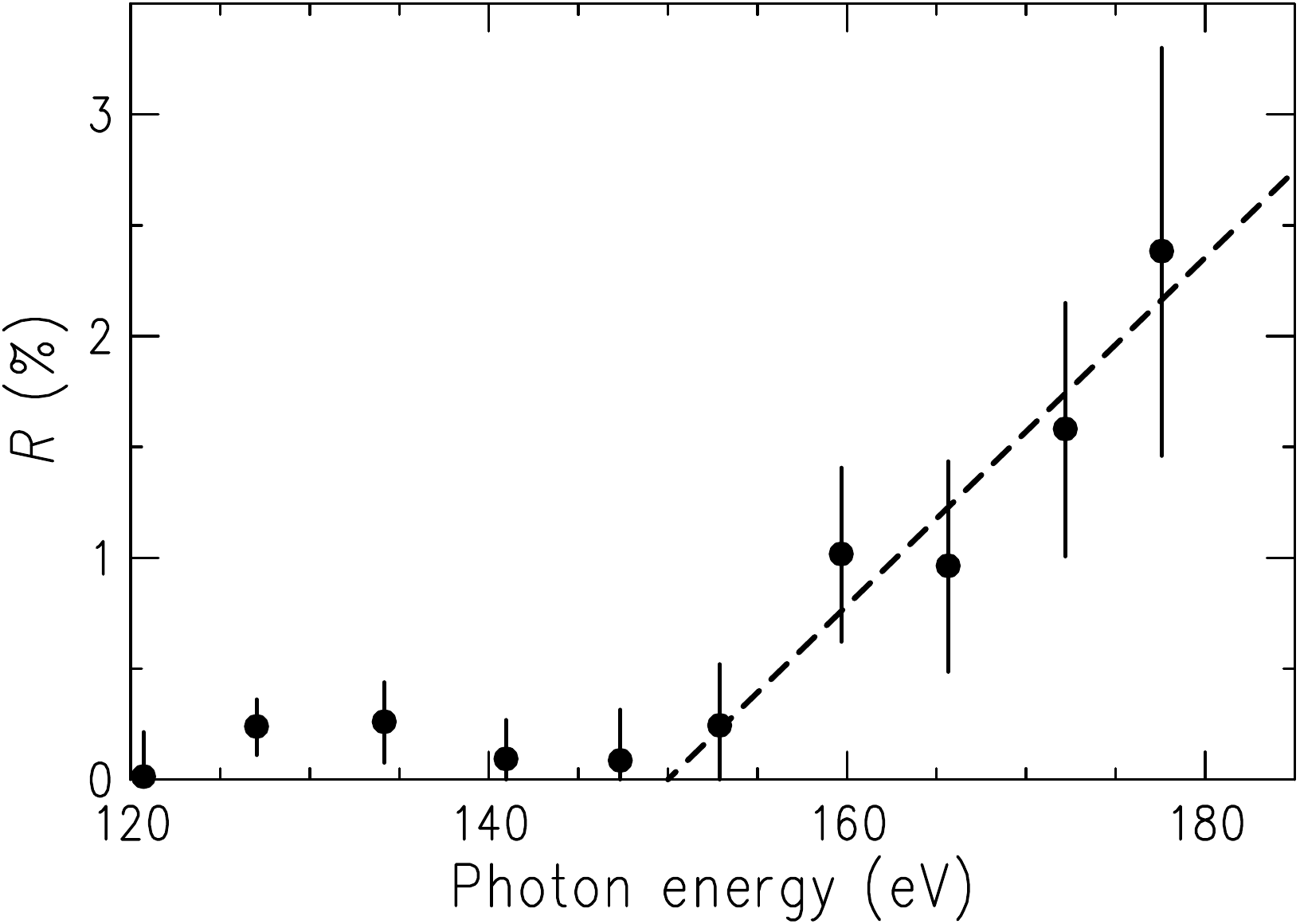}
\caption{\small High energy region of the DPI ratio of corannulene as 
defined in Eq.\ (\ref{one}) after subtracting the contribution of the 
knock-out mechanism. The dashed line is to guide the eye. 
}
\label{high}
\end{figure}

In a recent publication\cite{Wehl21}, the double photoionization ratio of 
corannulene  was compared to the DPI
ratio of coronene.  It was found that while both molecules had a  peak in 
the ratio near 50 eV, corannulene did not have the low energy peak that is  
discussed in Section \ref{coronene}.  Since corannulene has a ring of five 
interior carbon atoms, the question  arises as to why it has no peak, whereas 
coronene, with a ring of six interior atoms, has a peak at  10 eV.  In a 
recent analysis of the high-energy corannulene data we have found linear 
behavior in  the DPI ratio for photon energies above 150 eV, as displayed in 
Fig.\ \ref{high}. In this section, we will  show that the 5-carbon array 
in corannulene cannot form a closed loop.  Instead, there is an open  
array of carbon atoms that is the origin of the linear behavior in the DPI 
ratio.  

\begin{figure}[htb]
\includegraphics[width=4.0cm,clip]{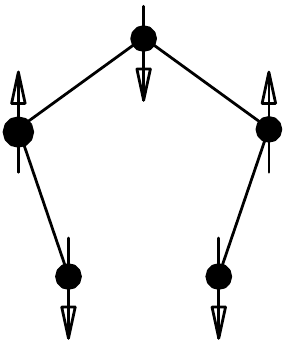}
\caption{\small Interior carbon atoms of corannulene with the spins
of their $\pi$ electrons.
}
\label{spins}
\end{figure}

The explanation for the difference in the DPI ratio between corannulene 
and coronene is related to the orientation of the spins of the interior 
$\pi$ electrons in their ground state which can be  either 
``up'' or ``down''.  In the coronene interior ground state, the six spins 
alternate between up and down and the total $\pi$ spin moment is zero.  In 
the corannulene case, where there are five interior $\pi$
electrons, the interior ground state is more complicated since there 
cannot be a complete  cancellation of the moments.  In Fig.\ \ref{spins}, 
we show an arrangement that minimizes the total  moment of the 5-spin 
interior array. 

The critical region in Fig.\ \ref{spins} is at the base of the pentagon, 
where the two down-spin $\pi$ electrons are nearest neighbors.  Because of 
the Exclusion Principle, it is impossible for a down-spin $\pi$ electron on  
one of the base sites to transfer to the other base site.  The effect of the 
Exclusion Principle is to introduce a 2-site ``impurity'' that 
eliminates the 5-site closed loop for $\pi$ electrons that have down spins.

DPI in corannulene involves the two carbon atoms for which the $\pi$ 
electrons have up-spins and the carbon atom on the top of the pentagon where 
there is a $\pi$ electron with a down-spin. The emission involves one of 
the up-spin electrons and the down-spin electron. In Ref.\
\onlinecite{Hube19} it is shown that when the DPI ratio is a linear 
function of the photon energy, the two electrons are emitted simultaneously 
with equal kinetic energy and opposite momenta. 

As a final comment, we note the similarity in the asymptotic behavior of 
the DPI ratios of pyrene and corannulene. In both cases, linear behavior 
was observed above 100 eV. We attribute the similarity to the arrangement 
of the interior carbon atoms. In the case of pyrene, there is a single 
carbon pair, whereas in corannulene, there are two interior pairing 
configurations that contribute to the DPI ratio

\section{SUMMARY}
\label{summary}
We have theoretically analyzed the resonance near 10 eV in the 
experimental DPI ratio of pyrene and coronene. Both molecules have 
interior as well as perimeter carbon atoms. Employing the Independent 
Subsystem Approximation, we can interpret the 10-eV resonance as 
being due to the interior carbon atoms. We also have presented an explanation
for the absence of the 10-eV peak in corannulene. 


\begin{acknowledgments}
We thank Dr.\ Pavle Jurani\'c for a critical reading of the manuscript.
This work is based upon research conducted at the Synchrotron Radiation
Center (SRC), University of 
Wisconsin--Madison, which was mainly supported by the UW Graduate School and
SRC Users.
\end{acknowledgments}



\end{document}